\documentclass[amssymb,prd,superscriptaddress,aps,nofootinbib,twocolumn,showpacs]{revtex4}
\usepackage{graphicx, epsfig, amssymb,bm} 
\usepackage{amsmath, amsfonts}
\usepackage{bm} 
\usepackage[breaklinks]{hyperref}
\usepackage{color}

\def\nn{\nonumber}
\def\be{\begin{equation}}
\def\ee{\end{equation}}
\def\beq{\begin{eqnarray}}
\def\eeq{\end{eqnarray}}
\def\pa{\partial}

\begin{document}

\title{Gravity with auxiliary fields}

\author{Paolo Pani}
\affiliation{CENTRA, Departamento de F\'{\i}sica, Instituto Superior
T\'ecnico, Universidade T\'ecnica de Lisboa - UTL, Avenida~Rovisco Pais 1, 1049
Lisboa, Portugal}
\affiliation{Institute for Theory $\&$ Computation, Harvard-Smithsonian
CfA, 60 Garden Street, Cambridge Massachusetts 02138, USA}

\author{Thomas P. Sotiriou} 
\affiliation{SISSA, Via Bonomea 265, 34136, Trieste, 
Italy {\rm and} INFN, Sezione di Trieste, Italy.}

\author{Daniele Vernieri} 
\affiliation{SISSA, Via Bonomea 265, 34136, Trieste, 
Italy {\rm and} INFN, Sezione di Trieste, Italy.}

\begin{abstract} 
Modifications of general relativity usually include extra dynamical degrees of freedom, which to date remain undetected. 
Here we explore the possibility of modifying Einstein's theory by adding solely nondynamical fields. With the minimal requirement that the theory satisfies the weak equivalence principle and admits a covariant Lagrangian formulation, we 
show that the field equations generically have to include higher-order derivatives of the matter fields. This has profound consequences for the viability of these theories. We develop a parametrization based on a derivative expansion and show that --~to next-to-leading order~-- all theories are described by just two parameters. Our approach can be used to put stringent, theory-independent constraints on such theories, as we demonstrate using the Newtonian limit as an example.
\end{abstract}


\maketitle

In four dimensions, the Lovelock theorem \cite{Lovelock:1971yv,Lovelock:1972vz} states that the only divergence-free rank-2 tensor which is constructed solely from the metric $g_{ab}$ and its derivatives up to second differential order is the Einstein tensor, $G_{ab}\equiv R_{ab}-\frac{1}{2}g_{ab}R$, plus a cosmological constant term (see also some previous, more restrictive proofs by Weyl~\cite{Weyl} and Cartan~\cite{Cartan}). This suggests a natural choice for the left-hand side of Einstein's equations (we work in units where $c=8\pi\,G=1$):
 \begin{equation}
  G_{ab}+\Lambda g_{ab} = \,T_{ab}\,. \label{Einstein}
 \end{equation}
The right-hand side, $T_{ab}$, is the matter stress-energy tensor and the contracted Bianchi identity then implies that $T_{ab}$ is divergence free, $\nabla_a T^{ab}=0$. This property is necessary for geodesic motion, which guarantees that the weak equivalence principle (universality of free fall) is satisfied. 

With the mild requirement that the field equations for the gravitational field and the matter fields be derived by an action, the arguments above single out the action
\begin{equation}
\label{action}
S=\frac{1}{2}\int d^4 x \sqrt{-g} R +S_{M}[g_{ab}, \psi]\,,
\end{equation}
where $\psi$ collectively denotes the matter fields, which couple minimally to $g_{ab}$, and so $S_M$ is understood to reduce to the Standard Model action in the local frame. 

Provided that one is not willing to give up the weak equivalence principle, Lovelock's theorem does not leave much room for modifying action (\ref{action}). The usual way to go around the theorem is to add extra dynamical fields. By doing so, however, one is inevitably introducing extra propagating degrees of freedom. Such degrees of freedom remain undetected to date. Hence, a major problem for alternative theories of gravity has been to tame the behavior of extra degrees of freedom, so as to evade current experimental constraints related to their existence~\cite{Will:2005va}. This comes in addition to the fact that it is not at all straightforward to construct theories with extra fields nonminimally coupled to gravity that avoid instabilities associated to the new degrees of freedom~\cite{Woodard:2006nt}.

What we wish to consider here is the much less explored option of adding \emph{nondynamical} extra fields. This is enough to circumvent Lovelock's theorem and, at the same time, it does not add extra degrees of freedom. The extra fields will then have to be auxiliary, {\em i.e.}~the field equations should allow them to be determined algebraically. A limited number of such theories, such as Palatini $f(R)$ gravities~\cite{Buchdahl:1983zz,Sotiriou:2008rp,DeFelice:2010aj} and Eddington-inspired Born--Infeld (EiBI) theory~\cite{Banados:2010ix}, have indeed been studied. Perhaps the most well-known example is Brans--Dicke theory with $\omega_0=-3/2$ (which is actually dynamically equivalent to Palatini $f(R)$ gravity~\cite{Sotiriou:2008rp,DeFelice:2010aj}).

One might be tempted to consider a modification of general relativity (GR) which is not restricted by Lovelock's theorem, as it refers to the right-hand side of the equations: more specifically, one may add any rank-2 tensor that is solely constructed by the metric and the matter fields and is identically divergence free, so as to not compromise the weak equivalence principle. However, it is unclear if such a tensor actually exists. Additionally it is reasonable to think that, if this theory is to come from an action, the corresponding modification to action~(\ref{action}) would amount to an addition of extra terms including the matter fields, hence introducing unacceptable modification to the equations of motion of the matter sector. 
 
Below we argue that, in theories that include auxiliary fields, eliminating them leads precisely (and generically) to a modification of Einstein's equation such as the one just described, without modifying the field equations of the matter fields. Using this fact, we then proceed to develop a very efficient parametrization of gravity theories with auxiliary fields. The power of this parametrization lies on the fact that it remains oblivious to the nature of the auxiliary fields and the way they enter in the action. As such, it allows one to study the phenomenology of such theories and impose observational constraints in a completely generic fashion. We finally use this parametrization to point out a specific important shortcoming of gravity theories with auxiliary fields (which has been pointed out for specific examples in Refs.~\cite{Barausse:2007pn,Barausse:2007ys,Barausse:2008nm,Sotiriou:2008dh,Pani:2012qd}): their equations contain higher-order derivatives of the matter fields and this  leads to nearly singular metrics in regimes where space-time is expected to be perfectly smooth.

Our argument is based on the following hypotheses: i) the theory admits a covariant Lagrangian; ii) in this formulation any extra field is auxiliary (see also below); and iii) the matter fields couple only to the metric in the usual way, so that the matter Lagrangian $L_M$ reduces to that of the Standard Model in the local frame. For concreteness we will focus on the simplest case of just one auxiliary field. However, this assumption is not crucial and the generalization to $N$ fields is straightforward. The theory is then described by the Lagrangian
%
\begin{equation}
 L=L_g[\mathbf{g},\bm{\phi}]+L_M[\mathbf{g},\mathbf{\psi}]\,;\label{lagr}
\end{equation}
$L_g$ describes the gravitational part where the metric $\mathbf{g}$ is possibly coupled to the extra field $\bm{\phi}$, for which the tensorial rank or other characteristics are left unspecified. 

Variation with respect to $\mathbf{g}$ and $\bm{\phi}$ yields
\begin{eqnarray}
 E_{ab}[\mathbf{g},\bm{\phi}]&=& T_{ab}\,, \label{eqG}\\
\mathbf{\Phi}[\mathbf{g}, \bm{\phi}] &=&0\,, \label{eqPhi}
\end{eqnarray}
where $E_{ab}$ is a generic rank-2 tensor. Variation with respect to $\psi$ will yield the field equations for the matter fields. 
Our requirement that $\bm{\phi}$ be an auxiliary field implies that, by using Eqs.~\eqref{eqG} and \eqref{eqPhi} (and possibly their derivatives) in some particular combination, it is possible to obtain an algebraic equation for $\bm{\phi}$, which can be \emph{schematically} written as
\begin{equation}
 {\cal F}[\bm{\phi},\mathbf{g},\mathbf{T}]=0\,, \label{algebraic}
\end{equation}
where $\bm{\phi}$ only appears at zeroth differential order.
Note that we do not necessarily require to solve Eq.~\eqref{algebraic} in \emph{closed} form. Indeed, it is sufficient to solve for $\bm{\phi}$ \emph{implicitly}, provided the implicit relation can be used to obtain a closed set of field equations for $\mathbf{g}$, where any dependence on $\bm{\phi}$ has been eliminated. 
It is clear that the matter fields will appear in Eq.~(\ref{algebraic}) only in the specific combination that forms the stress-energy tensor.

Let us assume that ${\cal F}$ does not depend on the matter fields at all. Then $\phi$ can be algebraically determined in terms of the metric only through ${\cal F}[\bm{\phi},\mathbf{g}]=0$. Consistency requires that Eq.~(\ref{eqPhi}) be trivially satisfied, and Eq.~(\ref{eqG}) reduces to
%
 $E_{ab}[\mathbf{g}]=T_{ab}$. But then, if $E_{ab}$ contains up to second derivatives of the metric, Lovelock's theorem requires $E_{ab}[\mathbf{g}]\equiv G_{ab}+\Lambda g_{ab}$ and the theory has to be GR. The case where $E_{ab}$ contains more than second derivatives of the metric does not concern us here, as the scope of adding auxiliary fields instead of dynamical ones was to avoid extra degrees of freedom.

On the other hand, if ${\cal F}$ depends on the matter fields, eliminating $\phi$ from Eq.~(\ref{eqG}) will yield
%
 $E_{ab}[\mathbf{g},\mathbf{T}]=T_{ab}$, 
which can be written without loss of generality as 
\begin{equation}
 G_{ab}+\Lambda g_{ab}=T_{ab}+S_{ab}[\mathbf{g},\mathbf{T}]\,. \label{modeom}
\end{equation}
The precise form of $S_{ab}$ will obviously depend on the specific form of the auxiliary field and how it enters the Lagrangian. However, $S_{ab}$ has to have the following properties: i) it vanishes when $T_{\mu\nu}=0$, as it was previously shown that when ${\cal F}$ is independent of the matter fields $E_{ab}=G_{ab}+\Lambda g_{ab}$; and ii) It is divergence-free, as a consequence of the contracted Bianchi identity and the fact that $T_{\mu\nu}$ is divergence free when the matter fields satisfy their field equations. Note that this latter property should hold identically, modulo the fact that $\nabla_a T^{ab}=0$, as it should not impose any further restriction to the dynamics. This is consistent with the fact that $S_{ab}$ came after eliminating an auxiliary field and that matter is minimally coupled to $\mathbf{g}$ in the Lagrangian~\eqref{lagr}. Lastly, it is worth mentioning that $\Lambda$ need not be identified with the cosmological constant that might appear in $L_g$.

We have shown that, in theories with an auxiliary field, eliminating the latter generically corresponds to modifying Einstein's equations by adding a divergence-free tensor that vanishes in vacuum. This tensor depends on the metric, the stress-energy tensor, and their derivatives. One could now set its origin aside and just attempt to construct it from its constituents. We proceed by doing so, order by order in the derivatives of the fields.  

The stress-energy tensor generically contains second derivatives\footnote{We will discuss  this hypothesis more extensively below.} (fermions being an exception) and this defines the lowest order. 
The only term one could add at the lowest order is $g_{ab} T$, where $T\equiv T^a{}_a$, so the equations can take the form
 \be
R_{ab} = T_{ab} -\alpha g_{ab} T +g_{ab} \Lambda +\ldots \,,\label{O1}
\ee
$\alpha$ being an arbitrary coefficient.
There are no terms with three derivatives one can construct. The terms with four derivatives are of three types: $\bm{T}^2$, ${\bm \nabla}^2 {\bm T}$, and contractions between ${\bm T}$ and the Riemann tensor. The only term that actually involves the Riemann tensor itself is $R_{(a|bc|d)} T^{bc}$, which can be eliminated without loss of generality in favor of other terms since $\nabla_c\,\nabla_d \,T^{ac} = R^a_{\phantom{a}bcd}T^{bc} + R_{bd}T^{ab}$.
Assuming that the perturbative expansion does not break down (which should be true at least in regimes where one expects to recover GR), one could use the lowest-order equation (\ref{O1}) in order to express $R_{ab}$ in terms of $T_{ab}$. Hence, up to fourth order in derivatives, we obtain
\beq
 S_{ab} &=& \alpha_1\, g_{ab}\,T   \label{eqexp} \nn \\
 &&+ \alpha_2\, g_{ab}\, T^2 + \alpha_3 \, T\, T_{ab} + \alpha_4\, g_{ab}\, T_{cd}\,T^{cd}    \nn \\
 &&+ \alpha_5 \, T^c\,_a\,T_{cb} + \beta_1\, \nabla_a\nabla_b\,T + \beta_2 \, g_{ab}\, \Box\,T   \nn \\
 &&+  \beta_3\, \Box\,T_{ab} + 2\beta_4\, \nabla^c\nabla_{(a}\,T_{b)c}+\ldots\,,
 \eeq
where $\alpha_i$ and $\beta_j$ are coefficients with appropriate dimensions. In the expression above we are not considering possible parity violating terms which would involve the Levi-Civita tensor.

We still need to impose that $S_{ab}$ be divergence free, at least to the required order, and this condition will impose some bond between the various coefficients. At first it might seem that the only solution is the trivial one, $\alpha_i=\beta_j=0$. However, this is not the case, as using the relations
\begin{eqnarray}
(\Box\,\nabla_b-\nabla_b\,\Box)\,T &=& R_{ab} \nabla^a\,T,   \label{rel1}\\
(\nabla^a\,\nabla^c\,\nabla_a-\nabla^c\,\Box)T_{cb} &=& R_{abcd}\nabla^d\,T^{ca},  \label{rel2}\\
\nabla^a R_{abcd} &=& 2\nabla_{[c}R_{d]b}\,,
\end{eqnarray}
and the lowest-order Eq.~(\ref{O1}) leads to cancellations between terms. Indeed, imposing $\nabla_a S^{ab}=0$ we obtain:
\beq
\alpha_1 &=& -\beta_1\,  \Lambda,  \,\,\,\,\,  4\alpha_2 = \left(1+2\alpha_1\right)(\beta_1-\beta_4), \nn \\
\alpha_3 &=& \beta_4\left(1+2\alpha_1\right)-\beta_1,  \,\,\,\,\,  2\alpha_4 = \,\beta_4\,,  \nn \\
\alpha_5 &=& - 2\beta_4\, ,  \,\,\,\,\,  \beta_2 = - \beta_1,  \,\,\,\,\,  \beta_3 = - \beta_4. \label{bonds}
\eeq 
The field equations finally read
\beq
 G_{ab} &=& T_{ab} - \Lambda g_{ab}   \label{eqexp2} \nn \\
 &-& \beta_1 \Lambda\, g_{ab}\,T + \frac{1}{4}\left(1-2\beta_1  \Lambda\right)(\beta_1-\beta_4)\, g_{ab}\, T^2    \nn \\
 &+&  \left[\beta_4\left(1-2\beta_1\, \Lambda\right) - \beta_1\right] \, T\, T_{ab} + \frac{1}{2}\,\beta_4 \, g_{ab}\, T_{cd}\, T^{cd} \nn \\
 &-& 2\beta_4\, T^c\,_a\,T_{cb} + \beta_1\, \nabla_a\nabla_b\,T - \beta_1\, g_{ab}\, \Box\,T  \nn \\
 &-&  \beta_4 \, \Box\,T_{ab} + 2\beta_4\, \nabla^c\nabla_{(a}\,T_{b)c} +\ldots\,,  
\eeq
where all coefficients are expressed in terms of $\beta_1$ and $\beta_4$.
It is worth noting that, although the equations above have been obtained as a derivative expansion, they could be equivalently obtained as a double expansion in the small-$\mathbf{T}$ and small-$\mathbf{\nabla T}$ limits. More precisely, introducing a further book-keeping parameter $\lambda$ associated to each derivative of the stress-energy tensor, it can be easily verified that Eq.~\eqref{eqexp2} is the most generic field equation which satisfies the aforementioned hypotheses to ${\cal O}(\mathbf{T}^2)$ and ${\cal O}(\lambda^2\mathbf{T})$. 
This equivalence hinges on the symmetries and on the tensorial rank of ${\mathbf T}$. Assigning a derivative order to $\mathbf{T}$ itself simply allows one to have a single book-keeping parameter and simplifies the discussion\footnote{Note also that, when one is working with an effective description of matter, such as fluids, quantities such as the energy density and pressure will not be of zeroth order even though they do not explicitly appear to contain derivatives. That can be understood intuitively by the fact that a scalar field admits an effective description as a perfect fluid.}.

Known theories with auxiliary fields do indeed fall within the parametrization developed above. EiBI gravity in the small coupling limit corresponds to $\beta_1=0$, $\beta_4=-\kappa/2$ using the definitions of Ref.~\cite{Pani:2012qd}.
Generic Palatini $f(R)$ theories correspond to $\beta_4=0$ with $\Lambda$ and $\beta_1$ being dependent on the model parameters. Interestingly, these two particular cases are in fact representative of two ``orthogonal'' classes of corrections. 

Our analysis demonstrates that theories with auxiliary fields, as well as any modification that does not allow for extra degrees of freedom, inevitably lead to equations with more than second derivatives of the matter fields. In the absence of extra dynamical fields, such terms have already been shown to be a major shortcoming in some specific theory: they lead to curvature singularities when there are sharp changes in the energy density of matter \cite{Barausse:2007pn,Barausse:2007ys,Barausse:2008nm,Sotiriou:2008dh,Pani:2012qd}. This problem will generically persist in the class of corrections we are discussing.

An analysis of the Newtonian limit is quite illuminating. For simplicity we set $\Lambda=0$. If the cosmological constant is to have the observed value then it can safely be considered as a higher post-Newtonian order contribution. 
In the limit of small velocities and small matter fields, one has $g_{ab}=\eta_{ab}+\epsilon h_{ab}$ and $T_{ab}=\epsilon \rho \delta_a^0\delta_b^0$, where $\epsilon\ll1$ is a book-keeping parameter.
We  define $\Psi_{ab}=h_{ab}-\frac{1}{2}\eta_{ab}h$, and indices are raised and lowered by the Minkowski metric $\eta_{ab}$. 
By performing the infinitesimal transformation $x^a\to x^a+\epsilon \xi^a$ where $\xi^a$ satisfies%
\begin{equation}
\label{gauge0}
 \epsilon \square \xi_b=\epsilon \pa_a\Psi^a_b-\zeta \pa_b T\,,
\end{equation}
we can impose a gauge such that $\epsilon \pa_a \Psi^a_b=\zeta \pa_b T$.
%
Here $\zeta$ is a numerical coefficient that we shall fix later on. To first order in $\epsilon$, and after some manipulations,  the field equations read $\nabla^2 h_{0i}=0$ and
\begin{eqnarray}
 -\frac{\nabla^2 h_{00}}{2}&=&\frac{\rho}{2}+\frac{1}{2}\beta_-\nabla^2\rho\,,\label{h00}\\
  -\frac{\nabla^2 h_{ij}}{2}&=&\frac{\delta_{ij}}{2}\left[{\rho}-\beta_+\nabla^2\rho\right]-\left[\beta_1-\zeta\right]\pa_i\pa_j\rho\,,
%
\end{eqnarray}
where $\beta_\pm=\beta_1\pm\beta_4$ and  $i,j=1,2,3$. It is now  evident that  setting $\zeta=\beta_1$ is the gauge choice that makes the spatial part of the metric diagonal~\cite{Will:2005va}. 
The solutions of the equations above then are $h_{0i}=0$ and
\begin{eqnarray}
 h_{00}&=&\int d^3x'\frac{\rho}{4\pi|\vec{x}-\vec{x}'|}-\beta_-\rho \,,\label{h00sol}\\
  h_{ij}&=&\delta_{ij}\int d^3x'\frac{\rho}{4\pi|\vec{x}-\vec{x}'|}+\beta_+\rho\,\delta_{ij} \,.\label{hijsol}
\end{eqnarray}
The last terms in Eqs.~(\ref{h00sol}) and (\ref{hijsol}) depend on the local value of the density. 
Thus, already at first order, it is evident that such an expansion does not fit into the standard parametrized post-Newtonian framework~\cite{Will:2005va}. The latter has to be extended to accommodate such corrections. It is also important to stress that the standard post-Newtonian expansion does not assume derivatives of the matter fields to be small. Therefore, a post-Newtonian expansion of Eq.~\eqref{eqexp2} would be valid at most to order ${\cal O}(T^2)$ and ${\cal O}(\lambda^2T)$.

If one considers the metric outside a spherical source, such as the Sun, these local  terms  vanish and the metric would be identical to the Newtonian metric in GR. This reflects the fact that {\em in vacuo} Eq.~(\ref{eqexp2}) reduces to Einstein's equation. However, when one considers what happens inside matter, and more specifically near the surface of an object, then the deviation from GR is drastic.

In particular, consider a situation where the density has a discontinuity, as can happen on the surface of a solid (or even for fluids that are described adequately by polytropic equations of state near the surface \cite{Barausse:2007ys}). Then the metric would be discontinuous there and the corresponding space-time singular. In fact, the gauge transformation~\eqref{gauge0} would not even be admissible, as the right-hand side would diverge and it would be impossible to eliminate the off-diagonal term of the metric. It is worth noting that this is not a coordinate problem, neither a problem associated with the Newtonian approximation. One can use Eq.~(\ref{eqexp2}) to straightforwardly calculate invariants such as $R$ or $R_{ab}R^{ab}$ and check that they diverge when $T$ is discontinuous, unlike GR.

One can argue that discontinuities in the density are not really physical and that the coarse-grained description of matter would break down, rendering our treatment inadequate. This is in principle true, but in practice it does not alleviate the problem. We know from everyday experience that very sharp transitions in density do exist in nature and one needs to go to very small scales to resolve them. This is enough to impose very tight constraints on $\beta_1$ and $\beta_4$. Let us demonstrate this with a simple example, a calculation of the acceleration $\vec{a}=\nabla h_{00}$ experienced within a thin layer in the interior and close to the surface of an object with Newtonian mass $M$ and radius $R_s$. The total acceleration reads
\begin{equation}
 \vec{a}=\vec{a}_N-\beta_-\nabla\rho\,,
\end{equation}
where $\vec{a}_N$ is the standard Newtonian acceleration. We assume spherical symmetry and for simplicity we take the density of the object to be nearly constant, $\rho(r)\sim\rho_0$, everywhere apart from a thin layer of width $L\ll R_s$ near the surface and the object is otherwise in vacuum.

If the thin layer were absent and the density had a jump, {\em e.g.}~$\rho(r)=\rho_0\Theta(R_s-r)$ where $\Theta$ is the Heaviside function, then the correction $\nabla\rho$ would introduce a Dirac delta contribution to the acceleration. 
This is already indicative of the pathology associated with having higher-order derivatives of matter in the gravitational field equations. 
Suppose now that microphysics in the transition region would allow for a smoother transitions that fails to be captured in the description above. Then one could consider the aforementioned layer to have the following density profile:
\begin{equation}
 \rho(r)=\rho_0\left[{(R_s-r)}/{L}\right]^n\,,  \qquad R_s-L<r<R_s\,.
\end{equation}
This can be thought of as an effective description for the smoother transition, where $L$ is the characteristic length scale at which microphysics would become important and $n$ parametrizes the slope of the profile. 
Using this profile, in the region $R_s-L<r<R_s$ we find:
\begin{equation}
 \frac{a}{a_N}=1+\frac{3n}{4\pi R_s L}\beta_-\left[{(R_s-r)}/{L}\right]^{n-1}\,,
\end{equation}
where we have used $a_N=M/R_s^2$, which is valid in the $L\ll R_s$ limit.
To not affect the standard Newtonian force to measurable levels in tabletop experiments, the last term on the right-hand side of the equation above must be much smaller than unity. Evaluating the acceleration at $r\sim R_s-L$ we obtain the constraint
\begin{equation}
 \left(\beta_1-\beta_4\right)\ll {4\pi R_s L}/{(3n)}\,.
\end{equation}
Note that, once $G$ is appropriately reinstated, each copy of $T_{ab}$ carries with it a $G$, as they only appear in this combination before eliminating the auxiliary field. Modulo fine-tuning, one could think of $\beta_1$ and $\beta_4$ as numerical coefficients of order unity times a characteristic length scale $\lambda_\beta$ squared. Then, choosing appropriate values for $R_s$ and $L$ one can turn the constraint above on a constraint on $\lambda_\beta$. For instance, if we choose --~quite conservatively~-- $R_s$ to be of the order of meters and $L$ to be of the order of microns, then
\begin{equation}
\lambda_\beta \ll n^{-1/2}\; {\rm mm} \,.
\end{equation}
Compared to typical astrophysical and cosmological length scales, this is an extremely tight constraint. 
For comparison, the Hubble radius squared is roughly $\Lambda^{-1}\sim 10^{52} {\rm m}^2$. 
One could hope to evade this constraint by fine-tuning the parameters. However, similar arguments can be made for the stresses $\sim\nabla h_{ij}$, which would provide a constraint on $\beta_+=\beta_1+\beta_4$ in Eq.~\eqref{hijsol}. Fine-tuning would not suffice to evade both constraints. 

It goes beyond the scope of our analysis to provide precise and exhaustive constraints on $\beta_1$ and $\beta_4$ (this will be done elsewhere). Our goal is to demonstrate that the theories we are discussing are unlikely to have any effect at large scales, if they are to be compatible with local experiments. Our analysis does not actually rule out the possibility that eliminating an auxiliary field can affect the value of the (effective) cosmological constant, which could perhaps be of some value in addressing the cosmological constant problem. However, the corresponding theory would have to accommodate at least two length scales apart from the Planck scale: $\lambda_\beta$ and the effective cosmological constant scale. Keeping these scales separated without fine-tuning would not be an easy task.

Our approach has certain limitations. $S_{ab}$ was constructed under the assumption that an expansion in derivatives (or, equivalently, a small-$\mathbf{T}$ and a small-$\mathbf{\nabla T}$ expansions) is applicable, which does not have to hold in all regimes. Our take on this is that such an expansion is expected to be valid in regimes where experiments verify the predictions of GR already. One could also wonder if going to the next order in derivatives could help relax the constraints. This is not the case, as adding more derivatives of ${\bm T}$ would simply make the metric even more sensitive to abrupt changes in the energy density. 

A subtle point is that coefficients of higher-order terms could ``contaminate" the relations of Eq.~(\ref{bonds}). This is due to the use of the lowest-order Eq.~\eqref{O1} in order to express $R_{ab}$ in terms of $T_{ab}$~--and specifically because of the presence of a cosmological constant. This is already seen in Eq.~(\ref{bonds}), where higher-order coefficients multiplied by $\Lambda$ are added to lower-order coefficients.  The presence of $\Lambda$ in such terms is required for reasons of dimensionality, and if $G$ were to be reinstated, all such terms would appear to be suppressed by $\Lambda\, G$. This guarantees that their contribution will be negligible. 

In summary, we have shown that gravity theories with auxiliary fields effectively lead to a modification of Einstein's equations by an addition of  a divergence-free second rank tensor which is constructed solely with the usual stress-energy tensor of matter, the metric, and their derivatives. 
It would be interesting to interpret these corrections as an effective stress-energy tensor~\cite{Delsate:2012ky}.
In these theories, the presence of higher-order derivatives of the matter fields is inevitable. We have developed a very general parametrization of auxiliary field theories and showed that, to next-to-leading order in derivatives of the matter fields, all auxiliary field theories can be described with only two parameters (apart from the cosmological constant). Finally, we have shown that these parameters can be severely constrained, as the presence of higher-order derivatives of the matter fields in the field equations renders the metric overly sensitive to abrupt changes of the matter energy density. This makes it particularly challenging to construct theories with auxiliary fields that could have any effect at large scales.

Other shortcomings, such as potential conflicts with the standard model \cite{Flanagan:2003rb} or issues with averaging \cite{Flanagan:2003rb,Li:2008fa},  have  been suggested  for Palatini $f(R)$ gravity, which is a specific auxiliary-field theory.
A more detailed analysis of the phenomenology of gravity theories with auxiliary fields, including a thorough discussion of these issues for generic theories, an extension of the parametrized post-Newtonian framework~\cite{Will:2005va}, and cosmological applications~\cite{Clifton:2011jh,Hu:2007pj}, will be presented in a forthcoming publication.


{\em Acknowledgments:}~T.P.S. and D.V. acknowledge financial support
from the European Research Council under the European Union's Seventh Framework Programme (FP7/2007-2013) / ERC Grant Agreement n.~306425 ``Challenging General Relativity'' and from 
the Marie Curie Career Integration Grant LIMITSOFGR-2011-TPS Grant Agreement n.~303537.
P.P. acknowledges financial support provided by FCT-Portugal through projects
PTDC/FIS/098025/2008, PTDC/FIS/098032/2008, CERN/FP/123593/2011 and by the European Community 
through the Intra-European Marie Curie contract aStronGR-2011-298297. 

\bibliographystyle{myutphys}
\bibliography{auxiliary}
\end{document}